\documentclass[preprint2]{aastex}

\shorttitle{Polarization measurements of GRB 041219A}
\shortauthors{G\"otz et al.}

\usepackage{natbib}

\begin{document}

\def\src{GRB 041219A}
\def\int{{\it INTEGRAL}}

\title{Variable polarization measured in the prompt emission of GRB 041219A using IBIS on board INTEGRAL}

\author{Diego G\"otz\altaffilmark{1}, Philippe Laurent\altaffilmark{2}, and Fran\c cois Lebrun\altaffilmark{2}}
\affil{CEA Saclay, DSM/Irfu/Service d'Astrophysique, F-91191, Gif sur Yvette, France}
\email{diego.gotz@cea.fr}

\author{Fr\'ed\'eric Daigne\altaffilmark{3}, and \v Zeljka Bo\v snjak}
\affil{Institut d'Astrophysique de Paris, UMR 7095
Universit\'{e} Pierre et Marie Curie-Paris 6 -- CNRS, 
98 bis boulevard Arago, 75014 Paris, France}

\altaffiltext{1}{Astrophysique Interactions Multi-\'echelles (AIM), CNRS}
\altaffiltext{2}{Astroparticules et Cosmologie (APC) - 10, rue Alice Domon et L\'eonie Duquet,  F-75205, Paris Cedex 13, France}
\altaffiltext{3}{Institut Universitaire de France}

\begin{abstract}

Polarization measurements provide direct insight into the nature of astrophysical processes. Unfortunately,
only a few instruments are available for this kind of measurements at $\gamma$-ray energies, and
the sources need to be very bright. Gamma-Ray Bursts (GRBs) are ideal candidates due to their large
flux over limited time intervals, maximizing the available signal-to-noise ratio. To date a few polarization
measurements have been reported, claiming of a high degree of polarization in the prompt emission of
GRBs but with low statistical evidence. 

We used the IBIS telescope on board the \int\ satellite 
to measure the polarization of the prompt $\gamma$-ray emission of the long and bright
\object{GRB 041219A}  in the 200--800 keV energy band. We find a variable degree of
polarization ranging from less than 4\% over the first peak to 43$\pm$25\% for the whole second peak.
Time resolved analysis of both peaks indicates a high degree of polarization, and the null average polarization in the first peak can be explained by the rapid variations observed  in the polarization angle and degree.

Our results are consistent with different models for the prompt emission of GRBs at these energies, but they favor synchrotron radiation from a relativistic outflow with a magnetic field which is coherent on an angular size comparable with the angular size of the emitting region ($\sim$1/$\Gamma$) . Indeed this model has the best capabilities to maintain a high polarization level, and to produce the observed  variability.
\end{abstract}

\keywords{gamma rays: bursts --- polarization --- gamma rays: observations}

\section{Introduction}

Gamma Ray Bursts (GRBs) are powerful flashes of $\gamma$-ray radiation appearing
at random directions on the sky, and lasting from a fraction to hundreds of seconds.
Some of them are firmly associated with supernovae of type Ib/c, and
the general picture is that bursts longer than 1 s are associated with the 
death of massive ($\gtrsim$ 30 M$_{\odot}$) stars (for a recent review on GRBs, see \citealt{meszaros06}).
The isotropic equivalent energy radiated during the prompt phase is of the order of $10^{51}$ to $10^{54} $ergs 
and is believed to originate from a highly relativistic outflow ($\Gamma \ga 100$) ejected by the central source.
The precise content of this jet, and especially its magnetization, as well as the details of the mechanism leading to
the $\gamma$-ray emission are still uncertain. Models range from unmagnetized fireballs where the observed
emission could be produced by relativistic electrons accelerated in internal shocks propagating within the outflow \citep{rees94}, to pure electromagnetic outflows where the radiated energy comes from magnetic dissipation \citep{lyutikov06}. Intermediate cases with mildly magnetized outflows are of course possible \citep[e.g.][]{spruit01}.
Even in the case of an unmagnetized fireball, a local magnetic field in the emission region, generated by the shocks, is necessary if the dominant process is synchrotron radiation from relativistic electrons.
In the case of a mildly magnetized outflow, such additional shock-generated magnetic field could also be present. The present situation is that neither the global magnetization of the outflow, nor the local intensity of the field are well constrained by the observations.

Polarization measurements of the prompt phase of GRBs can shed new light on the strength
and scale of magnetic fields, as well as on the radiative mechanisms at work.
Locally, the synchrotron emission has a high degree of linear polarization, that however
could be reduced by
relativistic effects in a jet. In the case where the magnetic field is mainly transverse and highly ordered, i.e. has a coherence scale which is larger than the typical size $\sim R / \Gamma$ of the visible part of the emitting region, the detected signal can still be highly polarized. On the other hand, in the case of a random field or an ordered magnetic field parallel to the expansion velocity, the polarization of the detected signal should vanish, except for the peculiar condition of a jet observed slightly off-axis \citep{waxman03}. 

To date only a few polarization measurements are available for GRBs. 
\citet{coburn03} 
reported a high degree of polarization, $\Pi = 80\% \pm 20\%$,
for GRB 021206.
However, successive reanalysis of the same data set could not confirm this claim, reporting
a degree of polarization compatible with zero \citep{rutledge04,wigger04}.
\citet{willis05} 
reported a strong polarization
signal in GRB 930131 ($\Pi > 35\%$) and GRB 960924 ($\Pi > 50\%$), but
this result could not be statistically constrained.
\src\ 
is the longest and brightest GRB localized by \int\ \citep{integral} to date \citep{vianello08}.
\cite{mcglynn07}, using the data of the \int\ spectrometer \citep[SPI;][]{spi},  reported a high degree
of polarization of the prompt emission 
($\Pi=68\pm29\%$) for the brightest part of this GRB. 

Here we report the polarization measurement of the prompt emission of \src, performed using
the imager on board \int, IBIS \citep{ibis}.

\section{Observations and analysis}
IBIS is composed of two position sensitive detector layers ISGRI \citep[CdTe, 15--1000 keV;][]{isgri}, and PICsIT \citep[CsI, 200 keV--10 MeV;][]{picsit}, and can be used as a Compton polarimeter \citep{lei97}, thanks to
the polarization dependency of the differential cross section for Compton scattering 
\begin{equation}
\frac{d\sigma}{d\Omega} = \frac{r_{0}^{2}}{2}\left(\frac{E^{\prime}}{E_{0}}\right)^{2}\left(\frac{E^{\prime}}{E_{0}}+\frac{E_{0}}{E^{\prime}}-2 \sin^{2}\theta \cos^{2}\phi \right)
\label{eq:cross}
\end{equation}
where $r_{0}^{2}$ is the classical electron radius, $E_{0}$ the energy of the incident photon, $E^{\prime}$
the energy of the scattered photon, $\theta$ the scattering angle, and $\phi$ the azimuthal angle relative
to the polarization direction. Linearly polarized photons scatter preferentially perpendicularly to the incident
polarization vector. Hence by examining the scatter angle distribution of the detected photons

\begin{equation}
N(\phi)=S[1+a_{0}\cos 2(\phi-\phi_{0})],
\label{eq:azimuth}
\end{equation}

one can derive the polarization angle, $PA =  \phi_{0} - \pi /2 + n \pi$, and the polarization fraction 
$\Pi= a_{0}/a_{100}$, where $a_{100}$ is the amplitude expected for a 100\% polarized source derived
by Montecarlo simulations \citep[see][]{forot08}.

\begin{figure}[ht]
\includegraphics[width=6cm, angle=-90]{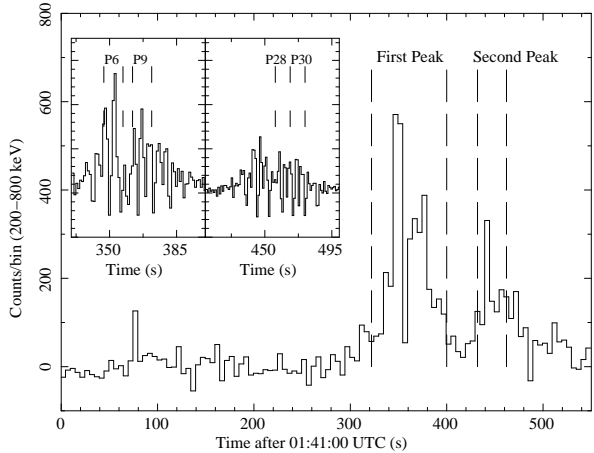}
\caption{Compton light curve of \src. Bin size is 5 s. The two insets show a magnified view of the two peaks, binned at 1 s. The analyzed intervals, are shown with dashed lines. P8 is omitted for clarity.}
\label{fig:comptonlc}
\end{figure}

IBIS is a coded mask telescope with a 29$^{\circ}\times$29$^{\circ}$ field of view at zero sensitivity, and
a 8.5$^{\circ}\times$8.5$^{\circ}$ central region where the sensitivity is maximal and uniform. 
\src\ was detected at $\sim$3.15$^{\circ}$ off-axis. Here we consider the events
which interacted once in the upper layer, ISGRI, and once in the lower layer, PICsIT, and whose
reconstructed energy lies in the 200 keV--800 keV range. These events are automatically selected on board
through a time coincidence algorithm, whose maximal allowed time window was 3.8 $\mu$s during
our observation. A light curve of the Compton events is reported in Fig. \ref{fig:comptonlc}.

\begin{figure}[ht]
\includegraphics[width=8cm]{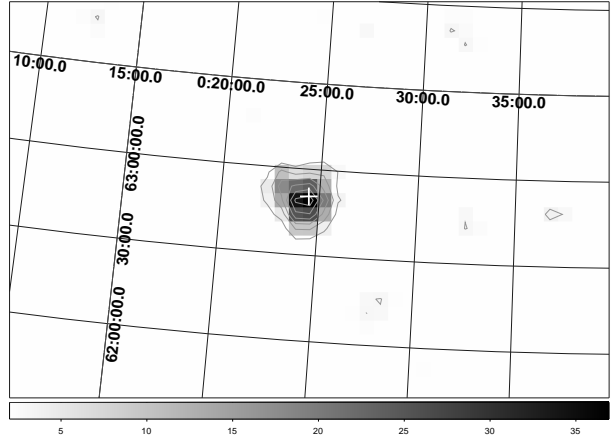}
\caption{200--800 keV Compton significance image of \src. The source is detected
at 37 sigma level. The white cross indicates the position of the optical counterpart \citep{vestrand05}, and grey contours are traced
at equal significance starting from 3 in steps of 6.}
\label{fig:comptonimg}
\end{figure}

As can be seen, the GRB is clearly detected, but due to its brightness, the available IBIS telemetry share was 
saturated, and hence not all the events could be sent to the ground. Images based on Compton events selected in the first and second peak show the bursting source respectively at a 32 sigma level and 20 sigma level.
An image made on the whole GRB is shown in Fig \ref{fig:comptonimg}, and the position derived ($\alpha_{J2000}$ = 00$^{h}$24$^{m}$24$^{s}$, $\delta_{J2000}$ = +62$^{\circ}$49$^{\prime}$44$^{\prime\prime}$ with an uncertainty of 1$^{\prime}$ at 90\% c.l.) is consistent with the one derived from simultaneous optical data \citep{vestrand05}.

To measure the polarization, we followed the same procedure described in \citet{forot08} that allowed to successfully detect a polarized signal from the Crab nebula. To derive the source flux as a function of $\phi$, the Compton photons were divided in 6 bins of 30$^{\circ}$ as a function of the azimuthal scattering angle. To improve the signal-to-noise ratio in each bin, we took advantage of the $\pi$-symmetry of the differential cross section (see Eq. \ref{eq:cross}), i.e. the first bin contains the photons with $0^{\circ}<\phi<30^{\circ}$ and $180^{\circ}<\phi<210^{\circ}$, etc.
Then the chance  coincidences (i.e. photons interacting in both detectors, but not related to a Compton event), have been subtracted from each detector image following the procedure described in \citet{forot08}. The derived detector images were then deconvolved to obtain sky images, where the flux of the source in each bin is measured by fitting the instrumental  PSF to the source peak. We finally fitted using a least squares technique the polarigrams (see Fig. \ref{fig:pola}) with Eq. \ref{eq:azimuth} to derive $a_{0}$ and $\phi_{0}$, and the errors on the parameters are dominated 
by the statistics of the data points. To evaluate the goodness of our fits, we computed the chance probability (see Eq. 2 in \citealt{forot08}) that our polarigrams are due to an unpolarized
signal, and reported these values in Fig. \ref{fig:pola}.

\section{Results}

\begin{figure}[ht]
\includegraphics[width=6cm, angle=90]{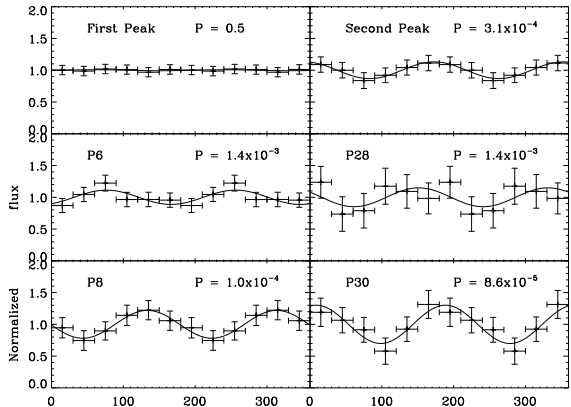}
\caption{Polarigrams of the different time intervals that have been analysed (see Table \ref{tab:pola}). For comparison purposes, the curves have been normalized to their average flux level. The crosses represent the data points (replicated once for clarity) and the continuous line the fit done using Eq. \ref{eq:azimuth}. The chance probability of a non-polarized signal is reported in each panel.}
\label{fig:pola}
\end{figure}

We analyzed the different portions of the GRB, focusing on the brightest parts.
First we analyzed the entire first and second peak, and then we performed a time resolved analysis:
36 intervals lasting 10 s, each one overlapping for 5 s with the previous one, have been analyzed over the whole duration of the GRB starting at 01:46:22 U.T. until 01:49:22 U.T. The most significant non-overlapping ones (P6, P8, P28, P30) have been chosen for the polarization analysis\footnote{Even if P9 is more significant than P8 (see Table \ref{tab:pola}), we prefer to discuss P8 because it is simultaneous to the
SPI analysis. Anyway, P8 and P9 results are statistically consistent.}. 
The integration times and the imaging
significance of the chosen time intervals are reported in Table \ref{tab:pola}, and Fig. \ref{fig:comptonlc}.
The azimuthal distributions of the GRB flux for the different time intervals are reported in Fig. \ref{fig:pola}. 

As can be seen, no polarization signal could be found integrating over the whole first peak, and the upper limit
is 4\%. On the other hand, a modulated signal is seen in the second peak corresponding to $\Pi = 43 \pm 25 \%$.
Integrating over smaller portions of the GRB, we measure
highly polarized signals, especially in P8, P9 and P30.

\begin{table}[ht]
\caption{Polarization results for the different time intervals.}

\begin{small}

\begin{center}
\begin{tabular}{cccccc}
\hline\hline
Name & T$_{start}$ & T$_{stop}$ & $\Pi$ & $PA$ & Image\\
& U.T.  & U.T. & \% & degrees & SNR \\
\hline
First Peak & 01:46:22 & 01:47:40 & $<$4 & -- & 32.0\\
Second Peak & 01:48:12 & 01:48:52 & 43$\pm$25 & 38$\pm$16 & 20.0\\
P6 & 01:46:47 & 01:46:57 & 22$\pm$ 13& 121$\pm$17 & 21.5\\
P8 & 01:46:57 & 01:27:07 &65$\pm$26 & 88$\pm$12 & 15.9\\ 
P9 & 01:47:02 & 01:47:12 &61$\pm$25 & 105$\pm$18 & 18.2\\
P28 & 01:48:37 & 01:48:47 &42$\pm$42 & 106$\pm$37 & 9.9\\
P30 & 01:48:47 & 01:48:57&90$\pm$36 & 54$\pm$11 & 11.8\\
\hline
\end{tabular}
\end{center}
Errors are given at 1 $\sigma$ c.l. for one parameter of interest.
\end{small}
\label{tab:pola}
\end{table}

\section{Discussion and conclusions}

Using SPI data \citet{mcglynn07} reported high polarization for \src,  but they restricted their analysis to the first peak of the GRB. They analyzed a 66 s interval within
the first peak, and a 12 s interval on the brightest part of it starting at 01:46:54 U.T. The latter corresponds to our P8 interval and indeed our results are compatible at 1 sigma level, both for the polarization fraction and angle, for which they find $\Pi$=68$\pm$29\% and 
$PA$=70$^{+14^{\circ}}_{-10^{\circ}}$ (100 keV--1 MeV). 
On the other hand we cannot confirm their result on the broader 
66 s time interval ($\Pi$=26$\pm$20\% and 
$PA$=70$^{+19^{\circ}}_{-27^{\circ}}$), starting at 01:46:21 U.T., which corresponds to our first peak analysis, where we do not
detect any polarized signal. By comparing P6 and P8, it seems that the polarization signal in this time interval results from the superposition of signals with different polarization angles, which could
give rise to a null signal over a longer time interval. The shorter duration of the second pulse,
where a polarization signal (P30) is clearly dominant over the others, could, on the other hand, explain the detection of 
a polarized signal over the whole second peak. Similarly, the 12 s peak emission may dominate the SPI
data, while in the IBIS data this time period is heavily affected by telemetry losses,
making the flux ratio between the brightest part (P8) and the rest of the GRB first peak smaller. In other words, the different polarization phases of the rest of the peak have a larger weight in our average measure.\\

The expected level of polarization of the prompt $\gamma$-ray emission in GRBs has been estimated by several authors
for different models, or variations within them. In most cases, the observed $\gamma$-ray emission
is due to the synchrotron radiation from relativistic electrons 
in the fast cooling regime. Their time-averaged distribution is a broken power law, $n(\gamma)\propto \gamma^{-p'}$ with $p'=p+1$ above $\Gamma_\mathrm{m}$ and $p'=2$ below, where $\Gamma_\mathrm{m}$ is the minimum Lorentz factor of the injected distribution of electrons, and $p\simeq 2-2.5$ its slope \citep{sari98}. The intrinsic polarization of the synchrotron radiation, $\Pi_\mathrm{syn}=(p'+1)/(p'+7/3)$ \citep{ribicky79} is then of the order of $\Pi_\mathrm{syn}=(p+2)/(p+10/3)\simeq 75\%$ above $\nu_\mathrm{m}$ and $\Pi_\mathrm{syn}=9/12\simeq 70 \%$ below, where $\nu_\mathrm{m}$, the peak of the spectrum in $\nu F_\nu$, is the synchrotron frequency of electrons at $\Gamma_\mathrm{m}$. High polarization levels can also be reached if inverse Compton scatterings are the dominant radiative process.\\ 

Our results show that (i) the polarization level in GRB04119A is varying on short time scales and can reach high values that correspond to a sizeable fraction of the intrinsic polarization $\Pi_\mathrm{syn}$. The polarization angle is varying as well, and (ii) the time-averaged value over longer intervals shows reduced polarization. We discuss these results in the context of different scenarios :\\
\noindent~(1) synchrotron emission from shock-accelerated electrons in a relativistic jet with an ordered magnetic field contained in the plane perpendicular to the jet velocity. This geometry is favored if the field is carried by the outflow from the central source, as the poloidal component decreases much faster with radius than the toroidal one. 
The polarization level at the peak of a given pulse can be as high as $\Pi/\Pi_\mathrm{syn}\sim 0.8$, i.e. $\Pi\sim 60 \%$, leading to a maximum time-averaged polarization in long intervals of $\Pi/\Pi_\mathrm{max}\sim 0.6 $, i.e. $\Pi\sim 45 \%$ in this case \citep{granot03a,granot03b,nakar03}. 
The main requirement is to have a uniform magnetic field in space, i.e. with a coherent scale $R \theta_\mathrm{B}$ with $\theta_\mathrm{B} \ga 1/\Gamma$. The fact that the polarization level and angle vary during the burst indicates on the other hand that the field is not necessarily 
uniform in time, as it would be required to explain a high level of the time-integrated polarization \citep{nakar03}. A magnetic field anchored in the central engine and carried by the outflow to large distance \citep[see e.g.][]{spruit01} is therefore not the only possibility. A magnetic field generated at the shock could also work, and even favor variability, if there is a process capable to increase the field coherence scale (the field is most probably initially generated on small, skin-depth, scales). The existence of such a process is unclear in our present knowledge of the micro-physics in mildly relativistic shocks \citep[see e.g.][]{keshet08}. Note that the condition $\theta_\mathrm{B} \ga 1/\Gamma$ is really necessary only in the pulses with the highest level of polarization. If $\theta_\mathrm{B}$ is smaller, so that a number $N\sim (\Gamma\theta_\mathrm{B})^{-2}$ of mutually incoherent patches are present is the visible region, the level of polarization will decrease, but the variability (both of the polarization level and angle) will increase \citep{granot03a}. If the radiating electrons are accelerated in internal shocks \citep{rees94,kobayashi97,daigne98}, the Lorentz factor is necessarily varying in the outflow, which can be an additional source of variability for the polarization. If $\theta_\mathrm{B}$ and $1/\Gamma$ are close, the number of coherent patches in the visible region could vary from a pulse to another. We therefore conclude that any scenario where the observed gamma-rays are produced by synchrotron radiation from electrons in a relativistic jet with an ordered magnetic field in the plane perpendicular to the jet velocity seems fully consistent with our observations, as long as the coherence scale $\theta_\mathrm{B}$ of the field is larger than $1/\Gamma$ in most of the emitting regions. A potential difficulty remains: an additional random component of the magnetic field is probably necessary to allow for particle acceleration in shocks. This component would
reduce the level of polarization by a factor that is however difficult to estimate, as the intensity of the the former is not well constrained \citep{granot03a,nakar03}; \\


\noindent~(2) synchrotron emission from a purely electromagnetic outflow. The estimated level of polarization is comparable with the previous scenario \citep{lyutikov03}. In addition, a magnetic field with a large coherence scale is naturally expected in such a purely electromagnetic outflow. One potential difficulty is, however, related to the mechanism responsible for the energy dissipation. It this scenario, the energy has to be extracted from the magnetic field before being radiated. Therefore magnetic dissipation must occur in the emitting region, changing the field geometry, which becomes  probably much less ordered, reducing the final level of polarization by a large factor \citep{lyutikov03,nakar03}. This effect is however difficult to estimate, as the details of the physical processes that could lead to
magnetic dissipation in such an outflow are still far from being understood; \\

\noindent~(3) synchrotron emission from shock-accelerated electrons in a relativistic jet with a random field generated at the shock and contained in the plane perpendicular to the jet velocity. A high level of polarization can be obtained even with a random magnetic field if the jet is observed from just outside its edge  \citep{ghisellini99,waxman03}.
The polarization at the peak of a given pulse can reach $\Pi/\Pi_\mathrm{syn}\simeq 0.8$, i.e. $\Pi \simeq 60 \%$ resulting in a time-integrated value of the order of $\Pi/\Pi_\mathrm{syn}\simeq 0.5-0.6$, i.e. $\Pi\simeq 40-45\%$ \citep{granot03a,granot03b,nakar03}. However these high values are obtained if the jet is seen with $\theta_\mathrm{obs}\simeq \theta_\mathrm{j}+1/\Gamma$, where $\theta_\mathrm{j}$ is the opening angle of the jet and $\theta_\mathrm{obs}$ the angle between the line-of-sight and the jet axis. Such viewing conditions are rare, except if $\theta_\mathrm{j}\sim 1/\Gamma$. 
Variability of the polarization level is expected if the Lorentz factor is varying in the outflow, as for instance in the internal shock model. 
Different pulses in the light curve correspond to viewing angles $\theta_\mathrm{obs}=\theta_\mathrm{j}+k/\Gamma$, where $k$
is larger for emitting regions with a larger Lorentz factor. 
The highest polarization is obtained in pulses with
$k\sim  1$ whereas the pulse flux decreases with $k$ for $k \ge 0$. 
The highest polarization
should therefore not be found in the brightest pulses, which is difficult to test as all pulses do not have necessarily the same intrinsic luminosity.
In addition, when several pulses superimpose, the measured polarization level, which is flux-weighted, could be reduced by a sizeable factor \citep{granot03b}.
Finally, the observed polarization can also be reduced if the jet edges are not sharp enough \citep{nakar03}. We conclude that this model cannot be
rejected as it can in principle reach high levels of polarization, but the conditions necessary for it seem more difficult to be achieved than in model (1). 
As shown by \citet{granot03b}, similar conditions as for scenario (3) would be required for a scenario where the field is ordered but parallel to the jet, leading to the same conclusions.\\


\noindent~(4) inverse Compton emission from relativistic electrons in a jet propagating within a photon field ("Compton drag" model). The level of polarization in this scenario can be even higher than for the synchrotron radiation and reach $60-100\%$, but only under the condition that the jet is narrow with $\Gamma\theta_\mathrm{j}\la 5$  \citep{lazzati04}. The maximum level of polarization is again obtained for $\theta_\mathrm{obs}\simeq \theta_\mathrm{j}+1/\Gamma$. These viewing conditions are very similar to those of models (3) and are discussed below in the case of GRB 041219A. Again, the polarization is reduced if the edges of the jet are not sharp enough. Variability of the Lorentz factor will again result in a varying polarization, with the same difficulties regarding the final level of polarization than in model (3). However, variations of the Lorentz factor could possibly be less large in this scenario as part of the variability of the light curve can be related to the inhomogeneity of the ambient photon field.\\

Combining the Yonetoku relation \citep{yonetoku04} to estimate the redshift and the isotropic energy, with the standard energy reservoir derived by \citet{frail01}, \citet{mcglynn07} have estimated that $\theta_\mathrm{j}\sim 2-3$ degrees in GRB 041219A, which leads to $\Gamma\theta_\mathrm{j}\sim (3.5-5) \left(\Gamma/100\right)$. Such an estimate, based on very debated relations, indicates that the peculiar viewing conditions necessary in the geometric models are difficult in the case of GRB 041219A but cannot be fully excluded. Models (3-4) can therefore not be rejected on the basis of these conditions only. As explained above, it is however not clear if a high polarization level can be maintained together with a high time variability. The status of model (2) is not clear as the details of the magnetic dissipation process are crucial and not well understood. Finally, our results favor model (1) 
as it seems to have the best capabilities to maintain a high polarization level with some variability. An additional random component of the magnetic field is however necessary for particle acceleration at the shock and its negative impact on the level of polarization should be carefully estimated. A final answer to distinguish between intrinsic and geometric models could be obtained by accumulating more observations. Indeed, models (1-2) predict a polarized emission for all bursts, whereas models (3-4) would predict that only a small fraction of GRBs are highly polarized. This idea has been tested recently by Montecarlo simulations performed by  \citet{toma08}, who conclude that if more than 30 \% of bursts are polarized, geometric models can be ruled out. This shows the importance of polarimetric measurements for the
understanding of intrinsic properties of GRBs, but the current instrumentation is statistically limited and can provide
measurements just for the brightest events. To significantly increase the sample of GRBs with measured polarization, future GRB mission with wide field of views and polarimetric capabilities are needed. 

\acknowledgements{D.G. acknowledges the French Space Agency (CNES) for financial support.
ISGRI has been realized and maintained in flight by CEA-Saclay/Irfu with
the support of CNES. Based on observations with INTEGRAL, an ESA project with instruments and science data centre funded by ESA member states
with the participation of Russia and the USA. }

\end{document}